\documentclass[a4paper, 10pt, twocolumn]{revtex4-1}
\usepackage{amsmath}
\usepackage{times}
\usepackage{graphicx}
\usepackage{setspace}
\usepackage[utf8]{inputenc}
\usepackage[T1]{fontenc}

 \usepackage{graphicx}
 \usepackage[usenames,dvipsnames]{xcolor}
 \RequirePackage{color}

 \definecolor{MyDarkGreen}{rgb}{0.02,0.60,0.06}

\begin{document}
%\singlespacing

\title{Network analysis of the {\emph{{\'{I}}slendinga s{\"{o}}gur}} -- the Sagas of Icelanders}
\author{P. Mac Carron \& R. Kenna}
\affiliation{Applied Mathematics Research Centre, Coventry University, Coventry, CV1 5FB, England}

\begin{abstract}
The {\emph{{\'{I}}slendinga s{\"{o}}gur}} -- or Sagas of Icelanders -- constitute a collection of medieval literature set in Iceland around the late 9th to early 11th centuries,  the so-called {\emph{Saga Age}}.
They purport to describe events during the period around the settlement of Iceland and the generations immediately following and constitute an important element of world literature thanks to their unique narrative style.
Although their historicity is  a matter of scholarly debate,
the narratives contain interwoven and overlapping  plots involving thousands of characters and interactions between them.
Here we perform a  network analysis of the {\emph{{\'{I}}slendinga s{\"{o}}gur}} in an attempt to gather quantitative information on interrelationships between characters and to compare {\emph{saga society}} to other social networks.
\end{abstract}

\maketitle

%%%%%%%%%%%%%%%%%%%%%%%%%%%%%%%%%%%%%%%%%%%%%%%%%%%%%%%%%%%%%%%%%%%%%%%%%%%%%%%%%%%%%%%%%
\section{Introduction}
\label{1}
%%%%%%%%%%%%%%%%%%%%%%%%%%%%%%%%%%%%%%%%%%%%%%%%%%%%%%%%%%%%%%%%%%%%%%%%%%%%%%%%%%%%%%%%%

The {\emph{{\'{I}}slendinga s{\"{o}}gur}}, or Sagas of Icelanders, are prose texts describing events purported to have occurred in Iceland in the period following its settlement 
in late 9th to the early 11th centuries.
It is generally believed that the texts were written in the 13th and 14th centuries by authors of unknown or uncertain identities but they may have oral prehistory.
The texts focus on family histories and genealogies and reflect struggles and conflicts amongst the early settlers of Iceland and their descendants.
The sagas describe many events in clear and plausible detail and
are considered to be amongst the gems of world literature and cultural inheritance.

In recent times, statistical physicists and complexity theorists have provided  quantitative insights into other disciplines, especially ones which exhibit collective phenomena emerging from large numbers of mutually interacting entities. 
In particular, a first application of network theory to the analysis of epic literature appeared in Ref.~\cite{EPL}. There, the  network structures underlying societies depicted in three iconic European epic narratives were compared to each other, as well as to real, imaginary and random networks.
A  survey of other multidisciplinary and interdisciplinary studies of complex networks is contained in Ref.~\cite{Costa2009}.
Different classes of networks have been identified according to various properties
in a manner akin to the classification of critical phenomena into universality classes in statistical physics~\cite{WattsStrogatz,Amaral}.
Such empirical studies have shown that social networks, in particular, usually have distinguishing properties;
they tend to be small world and are
well described by power-law degree distributions~\cite{Mislove},
they have high clustering coefficients~\cite{NewmanPark},  are often structurally balanced~\cite{Szell},
tend to be assortatively mixed by degree~\cite{Newman2002},
and exhibit community and hierarchical structures~\cite{Fortunato,Ravasz}.
While each of these properties is not unique to social networks, they are all commonly found in them and are hence characteristic of them.
The three 
epic narratives analysed in Ref.~\cite{EPL} exhibit some or all of these properties to varying degrees.

Here we report upon a network analysis of the {\emph{Sagas of  Icelanders}} -- the so-called {\emph{saga society}} \cite{Millar1990}.
Amongst ancient narratives, the {\emph{{\'{I}}slendinga s{\"{o}}gur}} present an especially interesting case study
because they purport to take place over a relatively short time period (namely around and following the settling of Iceland) and they contain an abundance of characters, many of whom appear in more than one narrative, allowing us to create a large, mostly geographically and temporally 
localised social network.
Our aim is to determine statistical properties of the saga society in a similar manner to
the studies of mythological networks in Ref.~\cite{EPL}, 
fictional network of Refs.~\cite{EPL,Alberich,Gleiser} and real social networks  in Ref.~\cite{Newman_review}, for example.
We also compare the social networks underlying the {\emph{{\'{I}}slendinga s{\"{o}}gur}} to each other and to random networks
to give unique insights into an important part of our cultural heritage.

In the next section, we present a brief overview of the {\emph{{\'{I}}slendinga s{\"{o}}gur}} to contextualise our report.
In Section~\ref{3} we gather the network-theoretic tools which are central to our approach.
The analysis itself is presented in Section~\ref{4} and we conclude in Section~\ref{5}.

%%%%%%%%%%%%%%%%%%%%%%%%%%%%%%%%%%%%%%%%%%%%%%%%%%%%%%%%%%%%%%%%%%%%%%%%%%%%%%%%%%%%%%%%%
\section{Sagas of the Icelanders}
\label{2}
%%%%%%%%%%%%%%%%%%%%%%%%%%%%%%%%%%%%%%%%%%%%%%%%%%%%%%%%%%%%%%%%%%%%%%%%%%%%%%%%%%%%%%%%%

\begin{table*}
  \begin{tabular*}{0.95\textwidth}{@{\extracolsep{\fill} }l|c c c c c c c c c c c c }
  \hline \hline
  Saga   & $N$  & $M$ &  $\langle{k}\rangle$ & $k_{\rm{max}}$ &   $\ell$ &  $\ell_{\rm{rand}}$ & $\ell_{\rm{max}}$ & $C$ &      $C_{\rm{rand}}$ & $S$ & $G_C$ & $\Delta$  \\
    \hline
%    \textbf{All} \\
    \small{{\emph{G{\'{i}}sla Saga}}}   &  103 & 254 & 4.9  &  44 & 3.4 & 2.9 & 11 & 0.6  &  0.05 & 10.8 & 98\%  & 9\%   \\
    \small{{\emph{Vatnsd{\ae}la Saga}}} &  132 & 290 & 4.4  & 31 & 3.9 & 3.3 & 10 & 0.5  &  0.03 & 12.7  & 97\%  & 2\%   \\
    \small{{\emph{Egils Saga}}}         &  293 & 769 &  5.3 & 59 & 4.2 & 3.4 & 12 & 0.6  &  0.02  & 25.5 & 97\%  & 5\% \\
    \small{{\emph{Laxd{\ae}la Saga}}}   &  332 & 894 &  5.4 & 45 & 5.0 & 3.5 & 16 & 0.5  &  0.02  & 19.0  & 99\%  & 6\%   \\
    \small{{\emph{Nj{\'{a}}ls Saga}}}   &  575 & 1612 & 5.6 & 83 & 5.1 & 3.7 & 24 & 0.4  &  0.01  & 31.0  & 100\% & 10\%  \\ \hline
    \small{Amalgamation of 5}                    & 1285 & 3720 & 5.8 & 83 & 5.2 & 4.1 & 16 & 0.5  &  0.005 & 80.4 & 99\%  & 7\% \\
    \small{Amalfamation of 18}                    & 1549 & 4266 & 5.5 & 83 & 5.7 & 4.3 & 19 & 0.5  &  0.004 & 98.7 & 99\%  & 7\% \\
    \hline \hline %\\
  \end{tabular*}
\caption{Properties of the full networks for  five  major sagas, for the amalgamation of the five sagas, and for the amalgamation
of all the data we gathered from 18 sagas.
Here $N$ and $M$ are the numbers of vertices and edges, respectively; $\langle{k}\rangle$ and $k_{\rm{max}}$ are the average and maximum degrees.
The average path length is $\ell$ and $\ell_{\rm{rand}}$ is the equivalent for a random network of the same size and average degree, while $\ell_{\rm{max}}$ is the longest shortest path.
The clustering coefficients  $C$ and $C_{\rm{rand}}$ are for the saga network and a random network of the same size and average degree. $S$ is the small world-ness, exceeding $1$ if the network is small world.
The size of the giant component as a percentage of the total network size is denoted by $G_C$
and $\Delta$ is the percentage of closed triads with an odd number of hostile links.}
\label{tab:Networks1}
\end{table*}

The {\emph{Sagas of Icelanders}} comprise an extensive corpus of medieval literature
``as epic as Homer, as deep in tragedy as Sophocles, as engagingly human as Shakespeare'' \cite{Smiley2000}.
We gathered data 
for 18 sagas and tales from the foundation of Icelandic %epic 
literature.
In addition to the longest and perhaps most famous {\emph{Nj{\'{a}}ls saga}} \cite{Njal},
we analysed the 17 narratives contained in \emph{The Sagas of Icelanders: a Selection} \cite{Smiley2000}.
There is considerable overlap between the various texts and the  18 selected narratives depict a sizeable proportion of the entire saga society.
Indeed, the combined saga society contained in the set of 18 tales comprises 1,549 individuals. Of these tales,
{\emph{Nj{\'{a}}ls saga}} (\emph{Njal's Saga}),
{\emph{Vatnsd{\ae}la saga}} (\emph{Saga of the People of Vatnsdal}),
{\emph{Laxd{\ae}la saga}} (\emph{Saga of the People of Laxardal}), 
{\emph{Egils saga Skallagr{\'{i}}mssonar}}  (\emph{Egil's Saga})
and
{\emph{G{\'{i}}sla saga S{\'{u}}rssonar}} (\emph{Gisli Sursson's Saga})
each have over 100 characters.
We  examine these individually in order to compare different types of saga and collectively to gain insight into the  structure of the overall saga social network.

%......................................................................
{\emph{Nj{\'{a}}ls saga}} is widely regarded as the greatest of the prose literature of Iceland in the Middle Ages and more vellum manuscripts  containing it have survived compared to 
any other saga~\cite{Rieu}. It also contains the largest saga-society network (see Table~\ref{tab:Networks1}).
The epic deals with blood feuds, recounting how minor slights in the society could escalate into major incidents and  bloodshed.
The events described are purported to take place between 960 and 1020~AD and, while most archaeologists believe the major occurrences described in the saga to be probably historically based,
there are clear elements of artistic embellishment.
%......................................................................
{\emph{Laxd{\ae}la Saga}}  tells of the  people of an area of western Iceland from the late 9th to the early 11th century. It has the second highest number of preserved
 medieval manuscripts and also contains the second largest network.
%......................................................................
{\emph{Vatnsd{\ae}la Saga}}
is essentially a family chronicle, following the settling of Ingimund, the grandson of a Norwegian chieftain,  in Iceland
with his family until the arrival of Christianity in the late 10th century.
%......................................................................
{\emph{Egils Saga}} tells of the exploits of a warrior-poet and adventurer.
The story begins in Norway with Egil's grandfather and his two sons. 
After one of them is killed, as a result of a dispute with the king, the family leaves to settle in Iceland. The latter part of the story is about the life of Egil himself.
%......................................................................
{\emph{G{\'{i}}sla Saga}} is an  
outlaw narrative 
centred on human struggles, as the eponymous character is
``on the run'' for 13 years before being finally killed. It is set in the period 940-980~AD. There are two versions of this and we use the version translated by 
Regal in Ref.~\cite{Smiley2000}.
%......................................................................

The latter story of an outlaw is mostly centred on one character rather than on a 
society and in this sense it is quite different to the other sagas considered here.
It is classed as an ``outlaw saga'' 
 as opposed to a ``family saga''.
{\emph{Egils Saga}} is also noteworthy in that a significant  proportion of it is set outside Iceland, 
beginning in Norway with the protagonist's family, where about a third of the saga's characters first appear.
Later in the story Egil travels to Norway amongst other places. Therefore the network contains overlapping social structures rather than a single coherent one. 
{\emph{Egils saga}}, moreover, contains a greater amount of 
supernatural elements than most of the sagas, though this is mostly contained in the prologue. 
{\emph{Egils saga}} is classed both as a ``poet's saga'' and a ``family saga''.
We will return to these observations in due course.

The narrative technique employed in the sagas is notable in that they are  objective in style.
Partly because of this, and the manner in which they are presented as chronicles, the 
sagas were widely accepted 
as giving more or less accurate and detailed accounts of early Icelandic 
life (obvious supernatural elements notwithstanding). 
The family sagas in particular, a corpus of almost 50 texts, are remarkable for their consistency.
As discussed in Ref.~\cite{Heather}, it is almost as if there is an ``unspoken consensus'' throughout the texts
concerning the make-up of the saga society: the main characters in one text appear as minor ones in another, giving the impression of an actual society.
More recently, however, historians have viewed the sagas more critically. While some  view the sagas as containing  romanticised but important elements of history, others  dismissed them as pure fiction, without any historical value.

An extensive discussion on the historical reliability of the various sagas
is contained in Ref.~\cite{Heather}.
It is suggested that they may be fiction framed in such a way as to appear historical to the modern reader.
However, even if the events are fictional, they may play out against a backdrop which includes real history.
In other words, the society may have been preserved in its essentials by oral tradition, while the events may be fictional.
Indeed, while it is also suggested 
that the society presented in such family sagas may be non-fictional,
it is lamented that it is ``almost impossibly difficult'' to distinguish fact from fiction in such sagas~\cite{Heather}.

Interpretative investigations such as that appearing in Ref.~\cite{Heather},
and indeed in 200 years of scholarly examination of the {\emph{{\'{I}}slendinga s{\"{o}}gur}},
tend to address questions surrounding events and individuals.
Here we focus instead upon the relationships between the characters depicted in the texts, the collection of which provides a spotlight onto the society depicted therein.
We present results from a network analysis of the {\emph{{\'{I}}slendinga s{\"{o}}gur}} and  show that the societal structure is similar to those of real social networks.

\begin{table*}
  \begin{tabular*}{0.95\textwidth}{@{\extracolsep{\fill} }l|c c c c c| c c c c c}
  \hline \hline
    Saga                               & \multicolumn{5}{c|}{Full network}   & \multicolumn{5}{c}{Friendly network} \\
                                       & $\gamma$       & $r_k$    &  $r_C$     & $n$ & $Q$   & $\gamma$ & $r_k$  &  $r_C$               & $n$        & $Q$  \\
    \hline
    \small{{\emph{G{\'{i}}sla saga}}} 
                      % {\red{(*outlaw*)}}
                      & 2.6(1)         & -0.15(5) & 0.01(7)    & 7  & 0.4 & 2.6(1)     & -0.14(5) & 0.02 (7)   &  9        & 0.5  \\
    \small{{\emph{Vatnsd{\ae}la saga}}}& 2.7(1)         &  0.00(6) & 0.08(6)    & 5  & 0.5 & 2.8(1)     &  0.00(6) & 0.10 (6)   &  5        & 0.6 \\
    \small{{\emph{Egils saga}}}  
                       %{\red{(*Norway*)}}
                       & 2.8(1)         & -0.07(3) & 0.28(4)    &  5  & 0.7 & 2.8(1)     & -0.03(4) & 0.35 (4)   &  6        & 0.7  \\
    \small{{\emph{Laxd{\ae}la saga}}}  & 2.9(1)         &  0.19(4) & 0.25(4)    & 12  & 0.6 & 2.9(1)     &  0.21(4) & 0.29 (4)   &  9        & 0.6  \\
    \small{{\emph{Nj{\'{a}}ls saga}}}  & 2.5(1) &  0.01(2) & 0.12(3)    & 12  & 0.3 & 2.6(1)     &  0.07(3) & 0.21 (3)   & 11        & 0.3  \\ \hline
    \small{Amalgamation of 5}              & 2.8(1)         &  0.05(2) & 0.17(2)    & 6  & 0.7 & 2.9(1)     &  0.09(2) & 0.23 (2)   &  8        & 0.6  \\
    \small{Amalgamation of 18}               & 2.9(1)         &  0.07(2) & 0.17(2)    & 9 & 0.7 & 2.9(1)     &  0.11(2) & 0.22 (2)   & 11        & 0.7  \\
    \hline \hline
  \end{tabular*}
\caption{The estimates for the exponent $\gamma$ for the various networks from fitting to Eq.(\ref{eqn:power}) and the assortativity coefficients $r_k$ and $r_C$  measured by degree and clustering. Here, $n$ is the number of communities when the modularity $Q$ reaches a plateau.}
\label{tab:Networks2}
\end{table*}

%%%%%%%%%%%%%%%%%%%%%%%%%%%%%%%%%%%%%%%%%%%%%%%%%%%
\section{Complex Networks}
\label{3}
%%%%%%%%%%%%%%%%%%%%%%%%%%%%%%%%%%%%%%%%%%%%%%%%%%%

In statistical physics, a social network is a graph in which vertices represent individuals, and edges represent interactions between them.
Edges are often undirected, reflecting a commutative nature of social interactions.
The {\emph{degree}} $k_i$ of an individual $i$ represents the number of edges linking that vertex to other nodes of the network. The average path length $\ell$ is the average number of edges separating two vertices.
The {\emph{clustering coefficient}} of vertex $i$ is given by
\begin{equation}
C_i = \frac{2 n_i}{k_i(k_i-1)},
\label{eqn:clustering}
\end{equation}
where $n_i$ is the number of edges linking the $k_i$ neighbours of  vertex~$i$~\cite{WattsStrogatz}.
In a social network, $C_i$ measures the proportion of an individual's acquaintances  who are mutually acquainted.
Topologically, it is the proportion of triads  having node $i$ as a vertex, which are closed by edges.
The {\emph{mean clustering coefficient}} $C$ of the entire network is obtained by averaging  Eq.~(\ref{eqn:clustering})
over all $N$ vertices.

A network is said to be {\emph{small world}} if its average path length $\ell$ is similar to that of a random graph
$\ell_{\rm{rand}}$ of the same size and average degree, and the average clustering coefficient of the network $C$ is much larger than that of the same random graph~$C_{\rm{rand}}$~\cite{WattsStrogatz}. A recent suggestion for a quantative determination of small world-ness is
\begin{equation}
S = \frac{C/C_{\rm{rand}}}{\ell/\ell_{\rm{rand}}},
\label{smallworldness}
\end{equation}
and the network is small world if $S>1$~\cite{Humphries}.

In keeping with our previous analysis \cite{EPL}, we may distinguish between friendly (positive) edges, in which the relationships may be characterised by  friendship, discussion, family connection, etc., and hostile (negative) edges, which involve %animosity or
physical conflict.
{\emph{Structural balance}}  is the tendency to disfavour triads with an odd number of hostile edges and
is related to the notion that `the enemy of an enemy is a friend' ~\cite{Heider,Cartwright}.
Examples of structural balance were recently found in systems as diverse as the international relations of nations \cite{Antal} and
the social network of a large-scale, multiplayer, online game~\cite{Szell}.

Denoting $p(k)$ as the probability that a given vertex has degree $k$, it has been found that the degree distribution for many complex networks follows a power law $p(k) \sim  k^{-\gamma}$ for a positive constant $\gamma$, so that
\begin{equation}
P(k) \sim  k^{1-\gamma},
\label{eqn:power}
\end{equation}
for the complementary cumulative distribution function~\cite{AlbertBarabasi}.

Eq.~(\ref{eqn:power}) usually starts at some minimum degree $k_{\rm{min}}$~\cite{Newman_Power}. 
If it were valid over the entire range of possible $k$-values, so that $k_{\rm{min}} = 1$, 
normalisation  would require $\gamma < 2$ and an expected mean degree which diverges.
However the average degrees of these networks are not large, 
therefore
we consider it reasonable to use $k_{\rm{min}}=2$.
This approach excludes peripheral nodes from the fit to Eq.(\ref{eqn:power}) only; it does not remove any nodes from the network itself.

%.....................................................................................
\begin{figure*}[t]
\begin{center}
\includegraphics[width=0.66\columnwidth, angle=0]{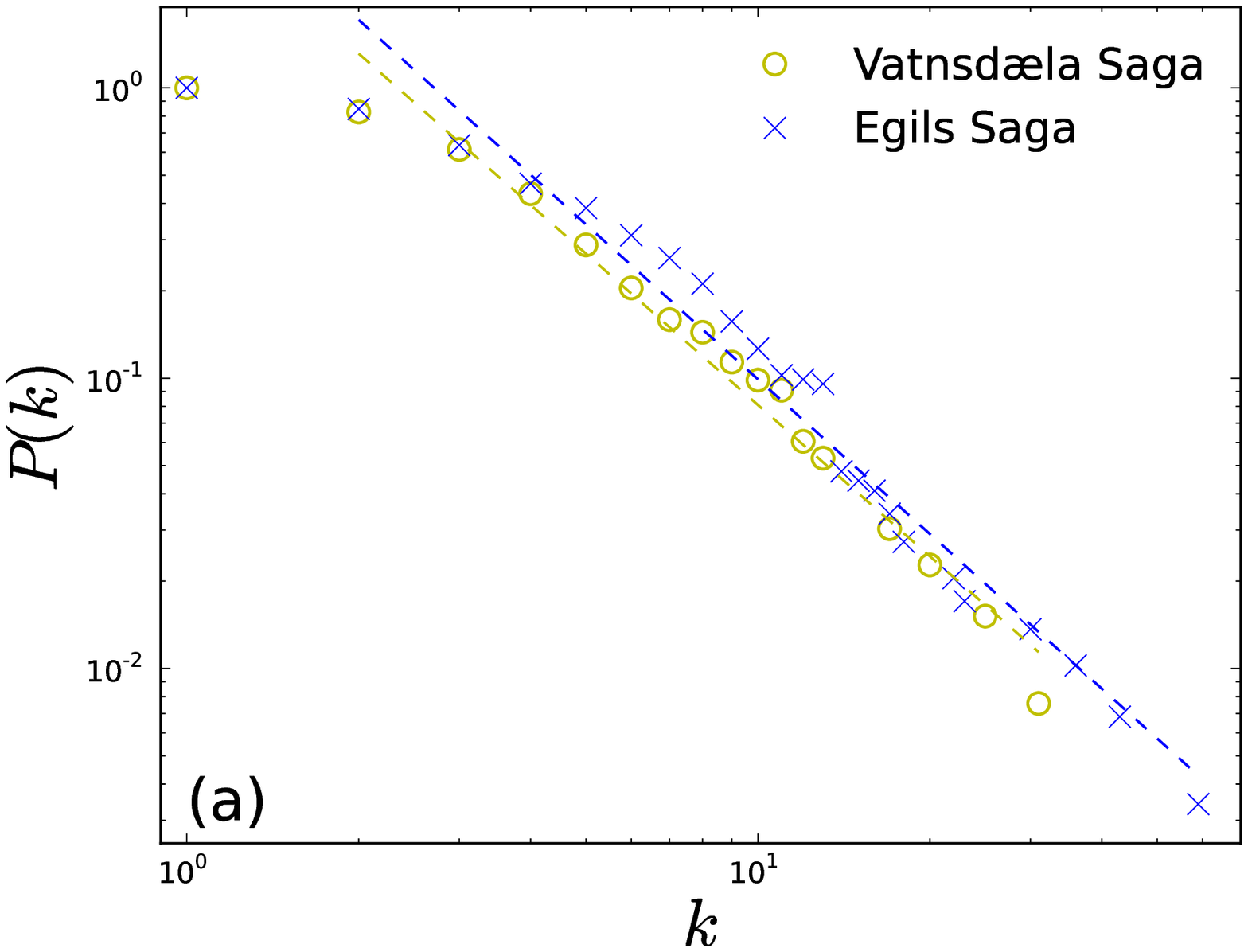}
\includegraphics[width=0.66\columnwidth, angle=0]{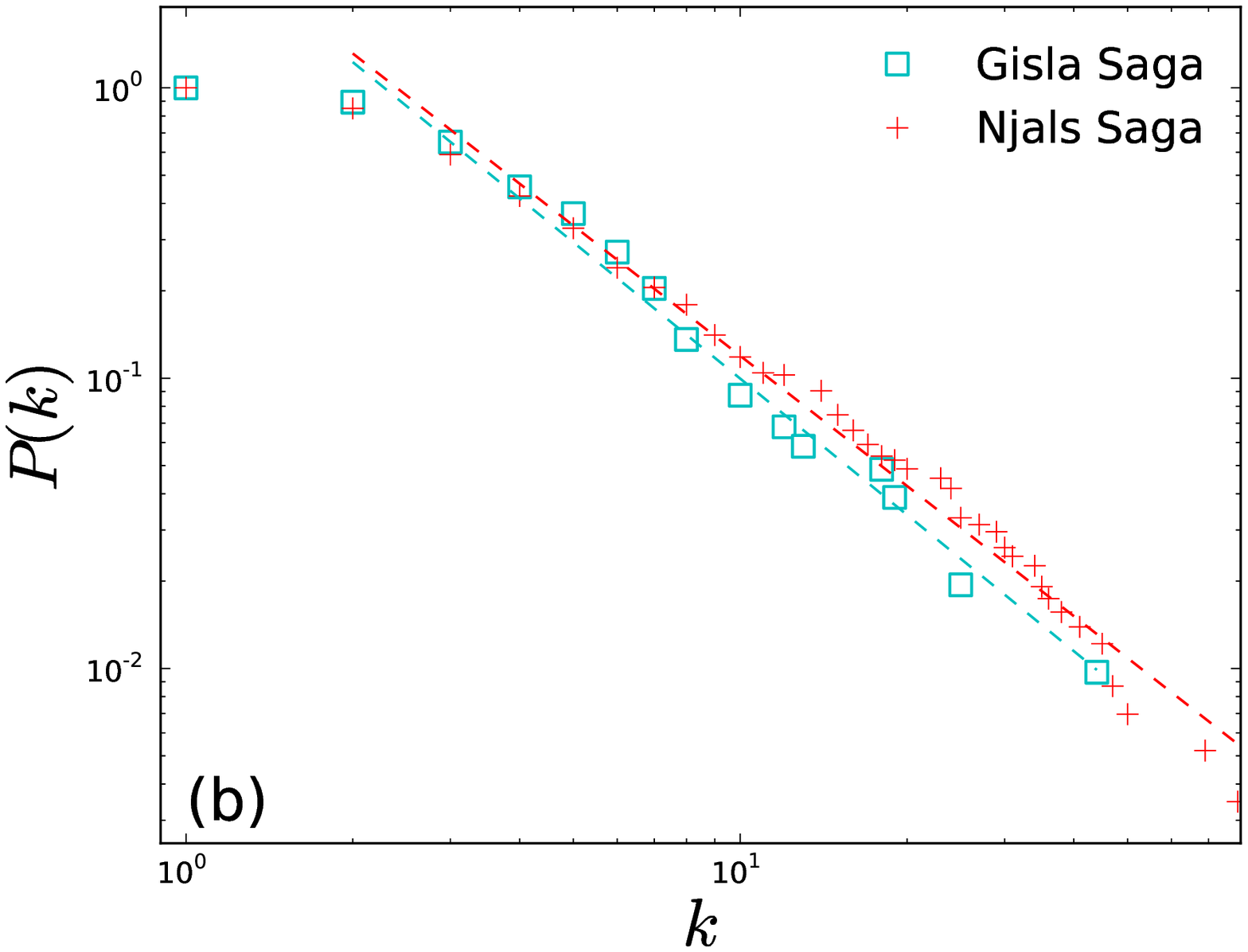}
\includegraphics[width=0.66\columnwidth, angle=0]{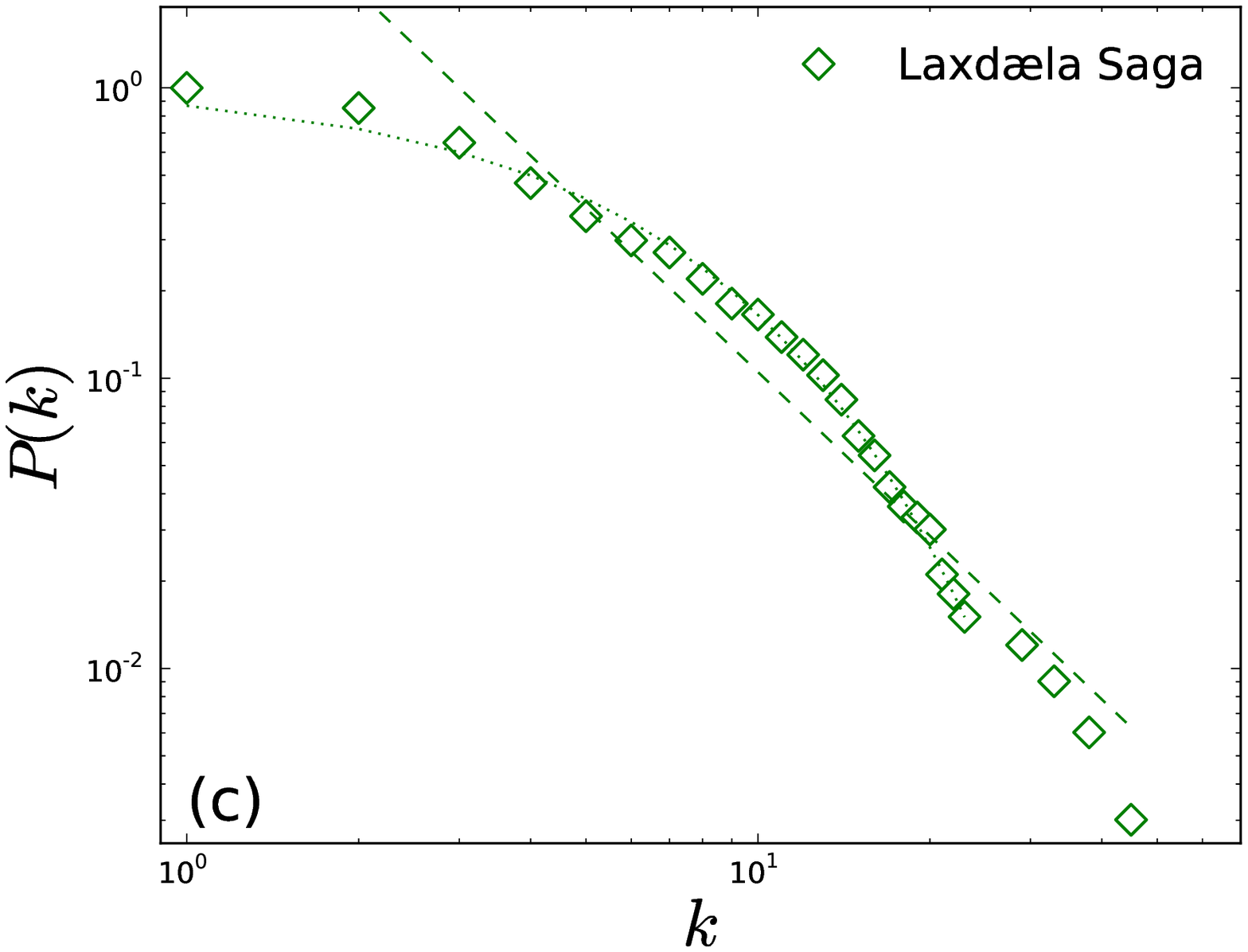}
\caption{The cumulative degree distributions for the five major sagas with power-law fits to Eq.(\ref{eqn:power}) using $k_{\rm{min}}=2$.
The degree-distribution descriptions of {\emph{Egils saga}} and  {\emph{Vatnsd{\ae}la saga}} in Panel (a) are similar, 
as are those for {\emph{Nj{\'{a}}ls Saga}} and {\emph{G{\'{i}}sla saga}} in Panel (b). 
Panel (c) shows {\emph{Laxd{\ae}la saga}} is different to the other four with the dashed and  
dotted lines representing power-law and exponential fits, respectively.}
\label{fig:degree_dist}
\end{center}
\end{figure*}
%.....................................................................................

The mean degree over the $N$ nodes of a network is defined as
\begin{equation}
 \langle{k}\rangle = \frac{1}{N} \sum_{i=1}^N{k_i},
 \label{eqn:average_k}
\end{equation}
and the second moment as
\begin{equation}
 \langle{k^2}\rangle = \frac{1}{N}\sum_{i=1}^N{k_i^2}.
\end{equation}
If $k_{e_1}$ and $k_{e_2}$ are the degrees of the two vertices at the extremities
of an edge $e$,
the mean degree of vertices at the end of an edge over the $M$ edges is
\begin{equation}
\bar{k} =  \frac{1}{2M}\sum_{e=1}^M (k_{e_1}+k_{e_2}),
\end{equation}
and 
the means  $\langle{ k }\rangle$ and $\bar{k}$ are related through % \cite{}
\begin{equation}
\bar{k} = \frac{\langle{k^2}\rangle}{ \langle{k}\rangle}.
\end{equation}
The {\emph{degree assortativity}} for the $M$ edges of an undirected graph is 
\begin{equation}
r_k = \frac{1}{M}\sum_{e=1}^M \frac{(k_{e_1}-\bar{k} )(k_{e_2}-\bar{k}) }{\sigma^2},
\label{assortativity}
\end{equation}
in which
\begin{eqnarray}
\sigma =  \left( \frac{1}{2M} \sum_{e=1}^{M} (k_{e_1} - \bar{k})^2 + (k_{e_2} - \bar{k})^2 \right)^{1/2} 
\end{eqnarray}
normalises $r_k$ to be between $-1$ and $1$~\cite{Newman2002}. 
If $ r_k < 0$ the network is said to be disassortative and if $r_k >0$ it is assortative.
One may also define the {\emph{clustering assortativity}} $r_C$ by replacing the node degrees $k_i$ by their clustering coefficients $C_i$ in Eq.(\ref{assortativity}). 
The  statistical errors can be calculated using the jackknife method~\cite{Efron,Newman2003}. 
It has been suggested that networks other than real social networks tend to be disassortatively mixed by degree~\cite{NewmanPark}. Real-world networks are also found to have high clustering assortativity $r_C$ and it has been suggested that this also indicates the presence of communities~\cite{Subelj}.

It has also been suggested that if the clustering $C_i$ decreases as a power of the degree $ k_i$, 
the network is {\emph{hierarchical}}~\cite{Ravasz}. 
In practise however, a power-law may not  describe the data well (see Refs.~\cite{Gleiser,Luduena} for example).
Nonetheless, a  decay signals that high degree vertices tend to have low clustering. 
In many sub-communities such nodes play an important role in keeping the entire network intact.

The {\emph{betweenness centrality}} of a given vertex is a measure of how many shortest paths (geodesics) pass through that node \cite{Freeman}. It therefore indicates how influential that node is in the sense that
vertices with a high betweenness control the flow of information between other vertices.
If $\sigma(i, j)$ is the number of geodesics between nodes $i$ and $j$, and if the number of these which pass through node $l$ is $\sigma_l(i, j)$, the betweenness centrality of vertex $l$ is
\begin{equation}
g_l = \frac{2}{(N-1)(N-2)}\sum_{i\neq j} \frac{\sigma_l(i,j)}{\sigma(i,j)}.
\label{eqn:betweenness}
\end{equation}
The  normalisation ensures that $g_l=1$ if all geodesics pass through node $l$.
An expression analogous to Eq.~(\ref{eqn:betweenness}) can be developed for edges to determine the {\emph{edge betweenness centrality}}.

Many social networks have been  found to contain community structure~\cite{Fortunato}. 
Here we follow Girvan and Newman  and  identify edges with the highest betweenness as these tend to connect such communities~\cite{GirvanNewman}.
Repeated removal of these edges breaks the network down into a number of smaller components $n$. 
To optimise  $n$, we investigate the {\emph{modularity}} $Q$~\cite{NewmanGirvan}.
We  define $E$ to be an $n \times n$ matrix, 
the elements $E_{ab}$ of which are the proportions of all edges in the full network that link nodes in community $a$ to nodes in community $b$. 
Denoting $F_a = \sum_b{E_{ab}}$, the modularity is then defined by
\begin{eqnarray}
Q = \sum_a (E_{aa} - F_a^2). 
\label{egn:Modularity}
\end{eqnarray}
If the structure comprises of only one community, such as typically the case for a random network,  $Q $ is close to zero. 
At the other extreme, if the network is  partitioned into $n$ sparcely inter-connected communities each containing approximately $M/n$ edges,
then $E_{ab} \approx \delta_{ab}/n $ and $F_b \approx E_{bb} \approx 1/n$, so that $Q \approx 1-1/n$.
Thus, although modularity is bounded by  $Q= 1$ for large $n$, it is typically between about $0.3$ and $0.7$ in social networks with varying degrees of community structure
\cite{NewmanGirvan}.

%%%%%%%%%%%%%%%%%%%%%%%%%%%%%%%%%%%%%%%%%%%%%%%%%%%%%%%%%%%%%%%%%%%%%%%%%%%%%%%%%%%%%%%%%%
\section{Networks Analysis of the Sagas}
\label{4}
%%%%%%%%%%%%%%%%%%%%%%%%%%%%%%%%%%%%%%%%%%%%%%%%%%%%%%%%%%%%%%%%%%%%%%%%%%%%%%%%%%%%%%%%%%

\begin{figure*}[t]
\begin{center}
\includegraphics[width=0.495\textwidth, angle=0]{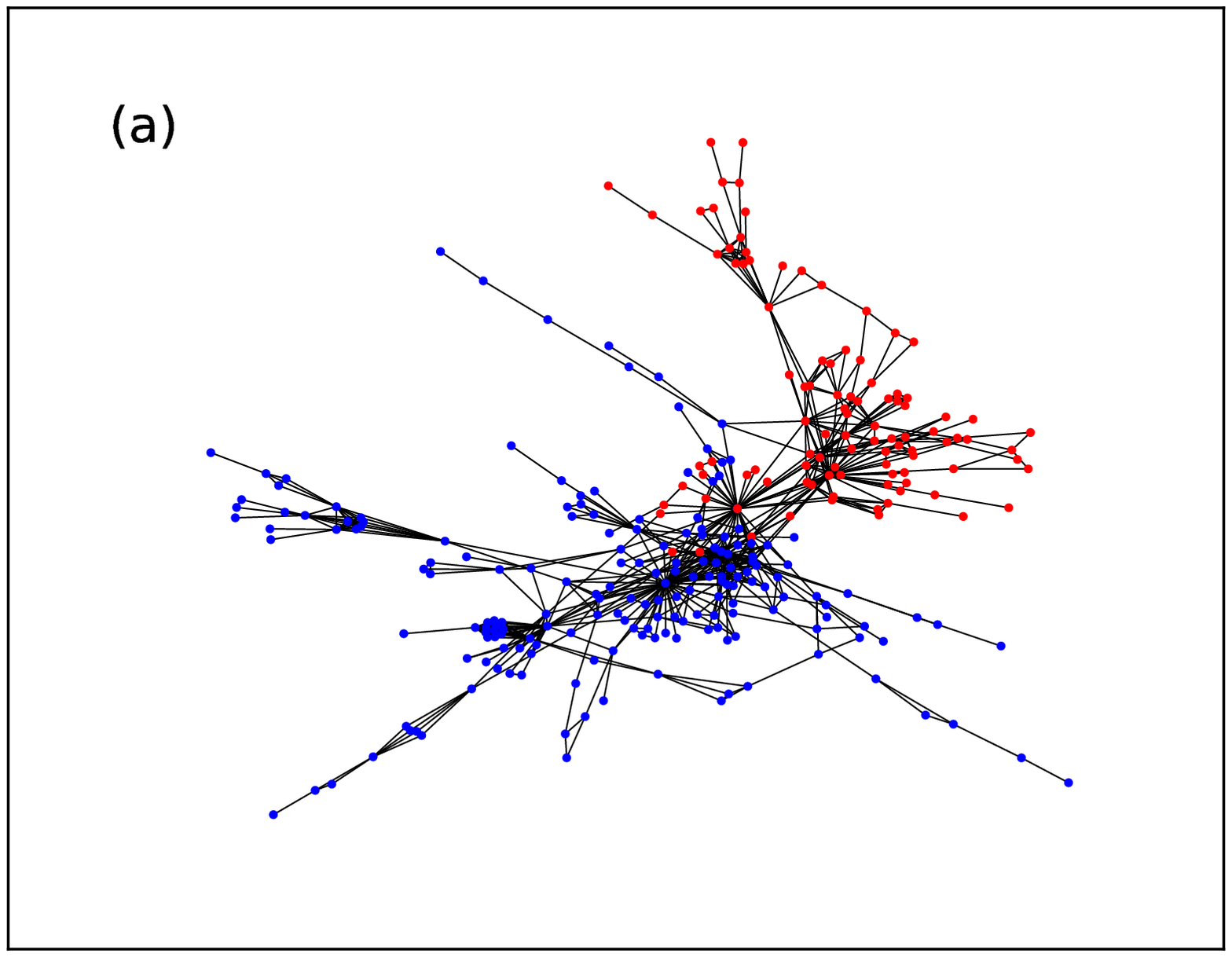}
%\hspace{-2cm}
\includegraphics[width=0.495\textwidth, angle=0]{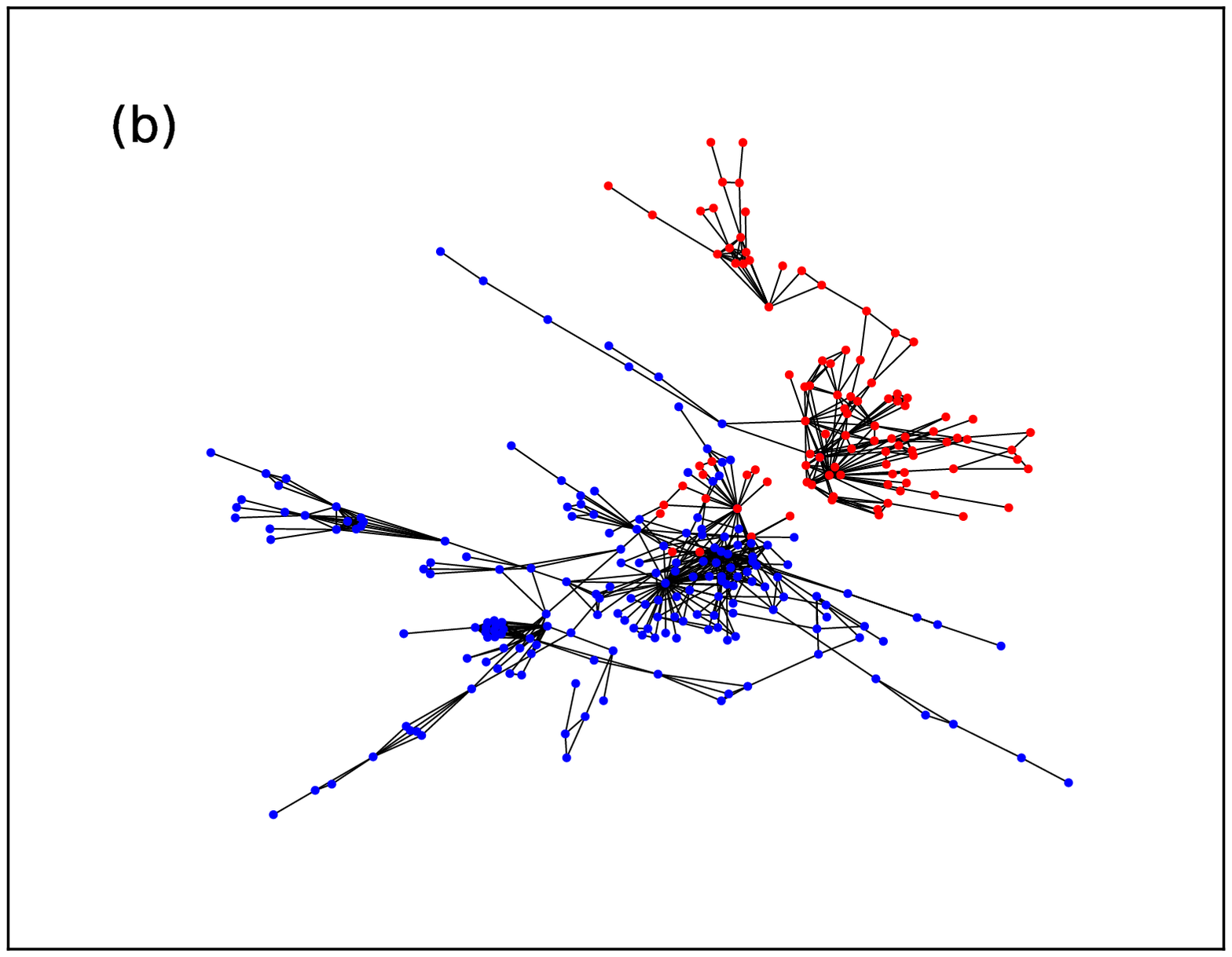}
%\vspace{-1cm}
\caption{(a) The network of {\emph{Egils saga}} drawn using the Fruchterman-Reingold algorithm~\cite{FruchtermanReingold}. 
Vertices coloured red represent characters who appear in the first part of the narrative while blue indicate those who appear during Egil's lifetime.
(b) The community algorithm successfully separates the two time periods despite some central characters being in both parts of the story.}
\label{fig:Egil}
\end{center}
\end{figure*}

In Tables~\ref{tab:Networks1} and~\ref{tab:Networks2}   the sagas are listed in order of network size and their properties tabulated for comparison.
The average degree $\langle{k}\rangle$ of the network can be calculated by Eq.~(\ref{eqn:average_k}) or simply by 2$M/N$, the factor $2$ entering because the networks are undirected. 
The longest geodesic (longest shortest path) is  $\ell_{\rm{max}}$.
In the table, $\ell_{\rm{rand}}$ and $C_{\rm{rand}}$ are the average path length and clustering coefficient
of a random network of the same size and average degree as that of the given saga or amalgamation of sagas.
The size of the giant component of the network as a percentage of its total size is  $G_C$.
The quantity $\Delta$ is the percentage of triads with an odd number of hostile links.
In Table~\ref{tab:Networks2}, the exponent $\gamma$  is estimated from a fit to the  power-law degree distribution 
(\ref{eqn:power}).

The average degree for each major network is similar, at about 5.
In each case, $\ell$ is comparable to, or slightly larger than $ \ell_{\rm{rand}}$, $C \gg C_{\rm{rand}}$ and $S>1$, as commonly found in social networks.
This is the {\emph{small-world}} property.
Each saga network has a giant connected component comprising  over $97\%$ of the characters in the networks which means that there are very few isolated characters in the societies as portrayed in these five sagas.

%STRUCTURAL BALANCE

The high clustering coefficients exhibited in each network signal large numbers of closed triads.
For each network,
under 10\% of these triads contain odd numbers of edges.
This means that odd numbers of hostile interactions are disfavoured and the saga societies are structurally balanced.

%%%%%%%%%%%%%%%%%%%%%%%%%%%%%%%%%%%%%%%%%%%%%%%%%%%%%%%%%%%%%%%%%%%%%%%%%%%%%%%%%%%%%%%%%%
\subsection{Individual major sagas}
%%%%%%%%%%%%%%%%%%%%%%%%%%%%%%%%%%%%%%%%%%%%%%%%%%%%%%%%%%%%%%%%%%%%%%%%%%%%%%%%%%%%%%%%%%

The complementary cumulative degree distributions for the five individual sagas are plotted in Fig.~\ref{fig:degree_dist}.
The $\gamma$ values for the various data sets fall in the range 
$ 2.6 
{\rm{\raisebox{-0.5ex}{ {\tiny  \shortstack{$<$ \\ $\sim$}} }}}
\gamma 
{\rm{\raisebox{-0.5ex}{ {\tiny  \shortstack{$<$ \\ $\sim$}} }}}
2.9$
with the full networks and the friendly subset giving  similar values.
However {\emph{Laxd{\ae}la saga}} is better fitted by an exponential distribution of the form $ p(k) \sim \exp{(-k/\kappa)}$, which delivers 
an estimate $\kappa = 5.5 \pm 0.1$. 
The power-law estimates indicate that the sagas' degree distributions are  comparable to each other and in the range $2 \le \gamma \le 3$ as usually found for social networks.

We next turn our attention to an analysis of  assortativity and community structures.  {\emph{Laxd{\ae}la saga}} has a  strongly assortative societal network.
{\emph{Nj{\'{a}}ls saga}} and {\emph{Vatnsd{\ae}la saga}}  are mildly assortative.
Only {\emph{G{\'{i}}sla saga S{\'{u}}rssonar}} is strongly disassortative like the small number of fictional networks so far analysed in the literature~\cite{EPL,Alberich,Gleiser}. As mention earlier however, this story is centred on a single protagonist's exploits instead of a larger society. 
This  is reflected in the fact that the protagonist's degree ($k=44$) 
is almost twice that of the next highest linked character ($k=25$) and may account for the high disassortativity.

{\emph{Egils saga Skallagr{\'{i}}mssonar}} appears mildly disassortative. As stated earlier, it is sometimes classed as a poet's saga instead of (or as well as) a family saga~\cite{Smiley2000}. It is interesting to note that the assortativity falls between that of the family sagas and an outlaw saga. It is
set in two different time periods, initially in Norway with the protagonist's father and grandfather, and then later with the life of Egil, beginning in Iceland and following his travels throughout his life.  
Therefore the network contains various social structures rather than a single cohesive one.
We can use community detection algorithms to investigate this.

Community structure is a prevalent feature of social networks.
We use Eq.~(\ref{eqn:betweenness}) to identify the edges with the highest betweenness and the Girvan-Newman algorithm~\cite{GirvanNewman} to break the networks down into smaller components, monitoring the modularity through Eq.~(\ref{egn:Modularity}).
The algorithm is halted when the modularity $Q$ first reaches a plateau. 
Community detection is particularly interesting for {\emph{Egils Saga}} as, unlike the other narratives, a significant portion occurs outside Iceland.
In Fig.~\ref{fig:Egil} the {\emph{Egils}} network is displayed, with red indicating characters which appear  before Egil's father leaves Norway 
and blue indicating characters who appear after this. 
Panel (a) displays  the entire {\emph{Egils}} network.
The modularity value reaches a plateau at $Q \approx 0.7$ with $n=5$ and severs 53 edges. 
As can be seen in Fig.~\ref{fig:Egil}(b), this  separates the two time periods. The sizes of the components are 112, 73, 61, 20 and 19.

The networks each exhibit clustering assortativity (signalled by $r_C>0$). 
This is also a common feature of real-world networks, both social and non-social~\cite{Subelj}.
In Ref.~\cite{Subelj}, it is also suggested that a high value of $r_C$ is a potential indicator for the presence of communities. {\emph{Egils saga}} and {\emph{Laxd{\ae}la saga}} indeed have high $r_C$-values and contain strong community structure.

\begin{figure}[t]
\begin{center}
\hspace{-3.6cm}
\includegraphics[width=0.975\columnwidth, angle=0]{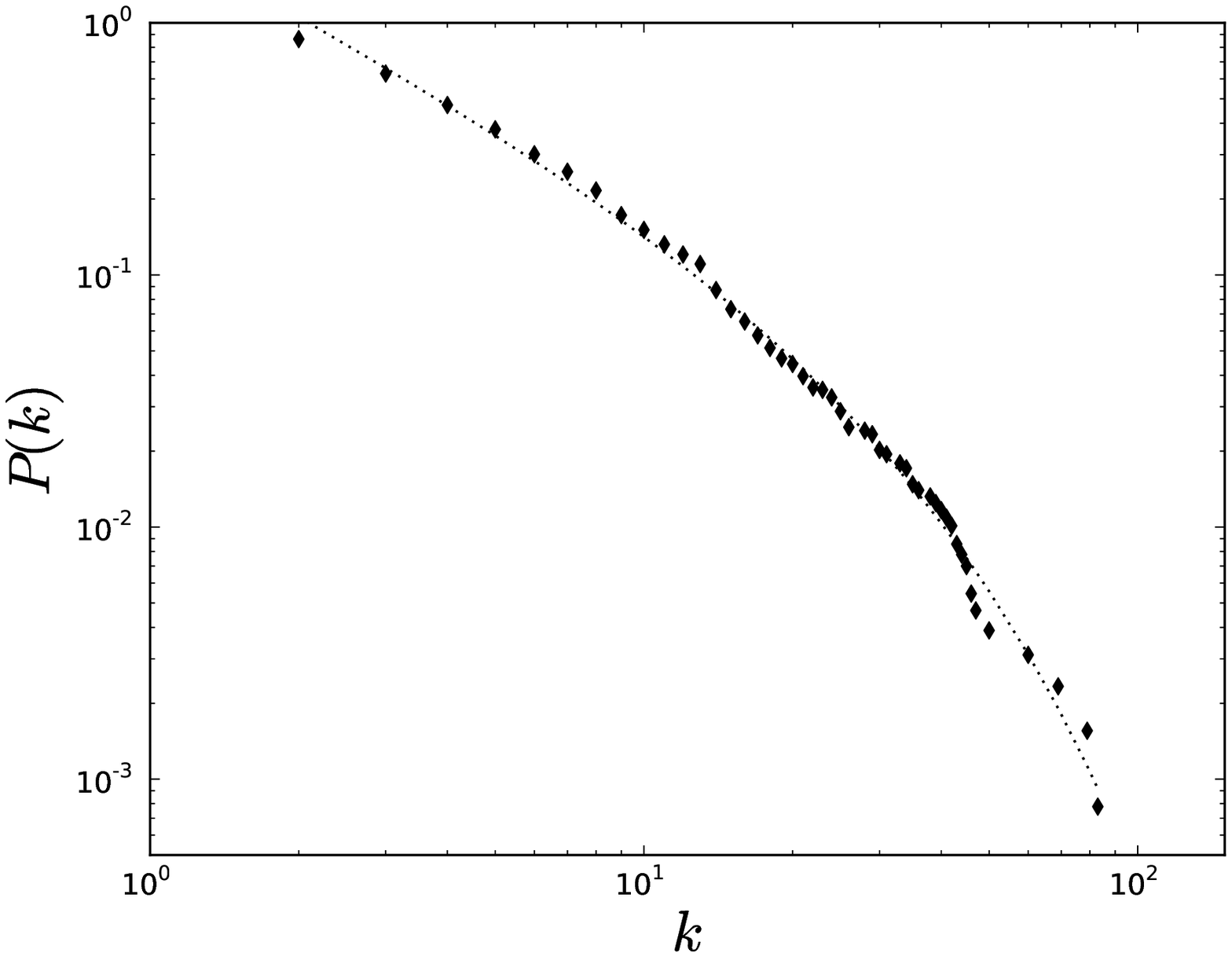}
\hspace{-3.6cm}
\llap{\raisebox{0.67cm}{\includegraphics[width=0.43\columnwidth, angle=0]{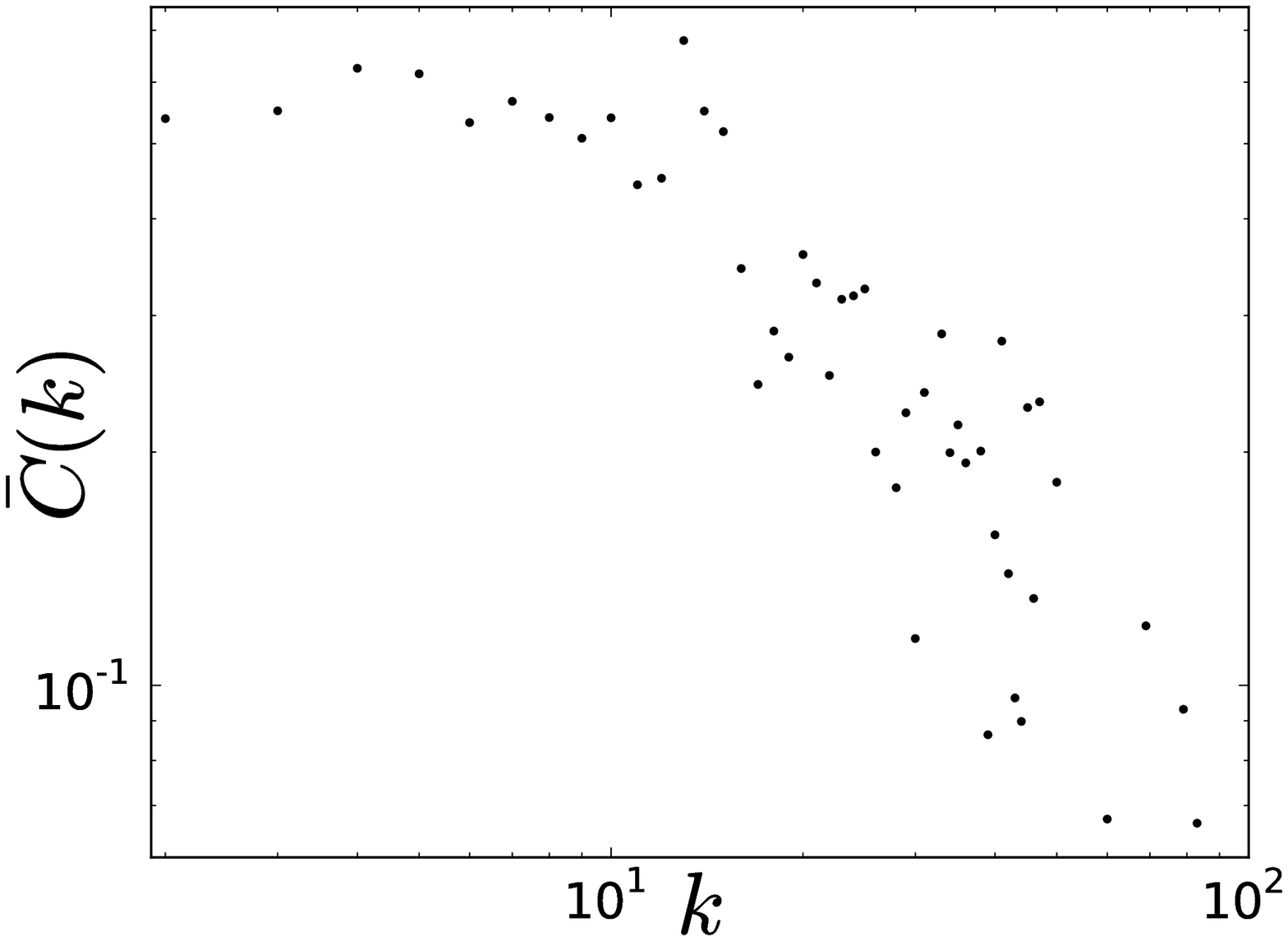}}}

\caption{The cumulative degree distribution (main panel) for the amalgamation of five major sagas. 
The dotted line represents a power-law fit with an exponential cut-off. 
The  mean clustering coefficient per degree is displayed in the inset panel for the same network.}
\label{fig:Five_sagas}
\end{center}
\end{figure}

In summary, the five individual major sagas have $\gamma$-values characteristic of 
many social networks studied in the literature and have varying degrees of assortativity. 
Although there are differences in detail between the networks of these 5 sagas, they also have many common 
features.

%%%%%%%%%%%%%%%%%%%%%%%%%%%%%%%%%%%%%%%%%%%%%%%%%%%%%%%%%%%%%%%%%%%%%%%%%%%%%%%%%%%%%%%%%%555
\subsection{Amalgamation of major sagas}
%%%%%%%%%%%%%%%%%%%%%%%%%%%%%%%%%%%%%%%%%%%%%%%%%%%%%%%%%%%%%%%%%%%%%%%%%%%%%%%%%%%%%%%%%%555

The network statistics for the amalgamation of the five major sagas are also given in Tables~\ref{tab:Networks1} and~\ref{tab:Networks2}.
This amalgamated network has 1,283 unique characters, only 15 of which do not appear in the giant connected component. Twelve per-cent of characters in the amalgamated network (152 characters in all)  appear in more than one of these major sagas.
Of these, 15 characters appear in three sagas and only one, Olaf Feilan, appears in 4, despite 
only having a degree of $k=13$ in total. No character makes an appearance in all five sagas.

Tables~\ref{tab:Networks1} and~\ref{tab:Networks2} indicate that the properties of the amalgamated network are similar to those of {\emph{Nj{\'{a}}ls saga}} individually. The average degree $\langle{k}\rangle$ increases reflecting the strong overlap between the characters in these sagas.
The amalgamated network also has similar properties to real social networks in that it is small world, structurally balanced, 
the degree distribution may be described by a power law with exponent between 2 and 3, and it is assortative.

The highest degree characters in {\emph{Nj{\'{a}}ls saga}} remain the highest degree characters in the amalgamated network. 
Moreover, unlike some of the lower degree characters, their degrees tend not to change on amalgamation as no new interactions involving them appear. 
Therefore the amalgamation process increases the proportion of low-degree characters over high-degree characters relative to {\emph{Nj{\'{a}}ls saga}},
leading to a faster decrease of the degree distribution  in the tail of Fig.~\ref{fig:Five_sagas}.
In fact, it is sometimes found for large networks that a cut-off for high degrees may be introduced in exponential form~\cite{Newman2001}, so that
\begin{equation}
P(k) \sim k^{1-\hat{\gamma}} e^{-k/\kappa}.
\label{eqn:truncated}
\end{equation}
In Fig.~\ref{fig:Five_sagas}(a), this fit is displayed with $\hat{\gamma} = 2.1 \pm 0.2$, $\kappa = 26 \pm 3$. 
Ref.~\cite{AlbertBarabasi} contains a list of exponents and  corresponding cut-offs for distributions of this type in a range of social networks.

The secondary, inset panel of Fig.~\ref{fig:Five_sagas} depicts the mean clustering per degree, $\bar{C}(k)$.
The decay may be interpreted as indicating  that high degree characters connect cliques, giving evidence of hierarchical structure~\cite{Gleiser,Luduena}.

%.....................................................................................
\begin{figure*}[t]
\begin{center}
\vspace{-2cm}\includegraphics[width=0.97\textwidth, angle=0]{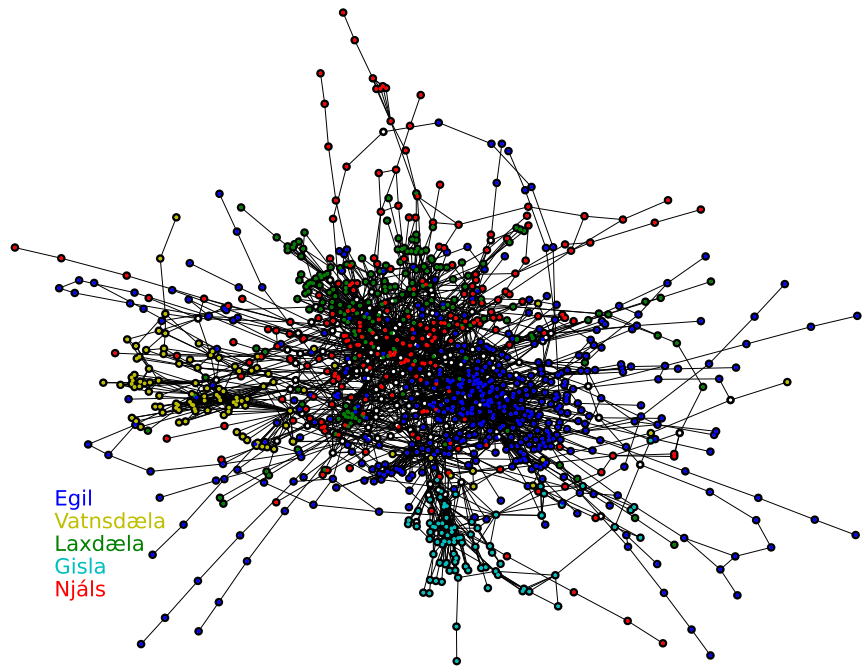}\vspace{-2cm}
\caption{Network for the amalgamation of the five major sagas (in colour online).
White nodes represent characters who appear in more than one saga. There is a large overlap of characters from {\emph{Laxd{\ae}la saga}} (green) and {\emph{Nj{\'{a}}ls saga}} (red). }
\label{fig:amalgamation}
\end{center}
\end{figure*}%.....................................................................................

We next break the amalgamated network back down to see if we can separate the  distinct sagas.
Again, we do this by removing the edges of highest betweenness.
This process gives an indication as to how interconnected the networks are.
Since there are five major sagas, we break the amalgamated network down until it has five large components.
These have sizes  670, 230, 136, 129, and 105, which are not  dissimilar to the sizes of each original network (Table~\ref{tab:Networks1}). 
Of the five emergent communities, those corresponding to {\emph{Egils saga}}, {\emph{Vatnsd{\ae}la saga}} and {\emph{G{\'{i}}sla saga}} emerge over 80\% intact -- see Table~\ref{table:components}.
However, the breakdown to five components delivers $Q \approx 0.5$ and fails to separate the societies of  {\emph{Nj{\'{a}}ls saga}} and {\emph{Laxd{\ae}la saga}} as, not only are there multiple characters that appear in both, but these characters often interact with different people in each narrative.

%...........................................................................
\begin{table}[b]
  \begin{tabular}{l l l}
  \hline
  Component & Main         & Secondary         \\ 
  size      & society      &        society\\ \hline
    670  & 67\% in {\emph{Nj{\'{a}}ls saga}} & 30\% in {\emph{Laxd{\ae}la Saga}} \\ % 451, 202, 81 Egil
    230 & 85\% in {\emph{Egils saga}} & 15\% in {\emph{Nj{\'{a}}ls Saga}} \\ %196, 36
    136 & 82\% in {\emph{Vatnsd{\ae}la saga}} &  13\% in {\emph{Nj{\'{a}}ls Saga}} \\ % 111, 18, 12 in Laxardal
    129 & 59\% in {\emph{Laxd{\ae}la saga}}& 51\% in {\emph{Nj{\'{a}}ls Saga}} \\ %77, 66
    105 & 85\% in {\emph{G{\'{i}}sla saga}} & 18\% in {\emph{Laxd{\ae}la Saga}}%89,18
  \end{tabular}
  \caption{Percentages of characters from different sagas which emerge when the amalgamated network is broken into 5 components. 
Note that the percentages can sum to more than 100 as the sagas share characters.}
  \label{table:components}
\end{table}
%...........................................................................

To separate {\emph{Nj{\'{a}}ls saga}} and {\emph{Laxd{\ae}la saga}}, one more step is required. 
Indeed, the modularity  for the full network reaches a plateau at $n=6$ communities with $Q \approx 0.7$.
the largest component now contains 463 characters, 91\% of which are from {\emph{Nj{\'{a}}ls saga}}.
The third largest component contains 207 characters, 80\% of which are from {\emph{Laxd{\ae}la saga}}.
However, the latter society emerges split into two separate components.

The large overlap between {\emph{Nj{\'{a}}ls saga}} and {\emph{Laxd{\ae}la saga}}
is visible in the network representation of  Fig.\ref{fig:amalgamation}.
In the figure, characters from each of the five major sagas are colour coded.
The characters in {\emph{Laxd{\ae}la saga}} appear the most scattered indicating that it is more weakly connected than some of the other sagas.  
This offers a potential new way to measure overlaps between  sagas, 
an issue which has been discussed in the literature~\cite{Rieu,Hamer}.

%%%%%%%%%%%%%%%%%%%%%%%%%%%%%%%%%%%%%%%%%%%%%%%%%%%%%%%%%%%%%%%%%%%%%%%%%%%
\subsection{Amalgamation of all 18 sagas}
%%%%%%%%%%%%%%%%%%%%%%%%%%%%%%%%%%%%%%%%%%%%%%%%%%%%%%%%%%%%%%%%%%%%%%%%%%%

Finally we amalgamate all 18 sagas and tales for which data were harvested.
The  statistical properties are again given in Tables~\ref{tab:Networks1} and~\ref{tab:Networks2}.
When all 18 sagas are amalgamated $\langle{k}\rangle$ decreases slightly relative to the corresponding value for the amalgamation of only the five major sagas, signalling that there are numerous low degree characters added to the network.
For the amalgamation of 18 sagas, the network is again small world, structurally balanced, hierarchical and assortative.
The giant component contains 98.6\% of the 1,547 unique characters.

\begin{figure}[!t]
\begin{center}
\includegraphics[width=0.9\columnwidth, angle=0]{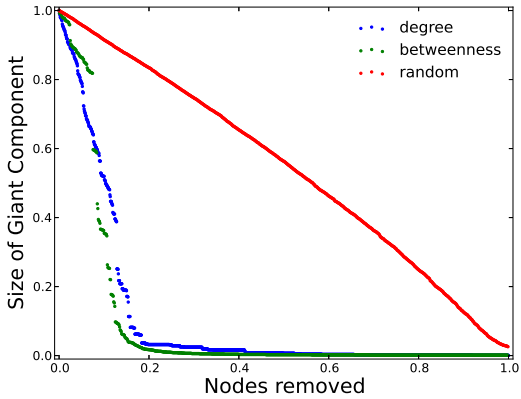}
\caption{The size of the giant component as a fraction plotted against the number of nodes removed as a fraction. It is least robust when attacked by betweenness.}
\label{fig:Robustness}
\end{center}
\end{figure}

Complex networks are often found to be robust to random removal of nodes but fragile to targetted removal (for an overview of network resilience see \cite{AlbertBarabasi,Newman_review}).
To test its robustness we remove characters starting with those of highest degree or betweenness. We also remove characters randomly.
In the latter case, we report the average effects of 30 realisations of random removal of nodes.
Like other social networks the amalgamation is robust when nodes are randomly removed; removing 10\% of the nodes (155 characters) leaves the giant component with  94\% of the characters in the network on average.
Removing the characters with the highest degrees  or highest betweenness centralities causes the network to breakdown more rapidly; removing 10\% brings the giant component to about half its original size.
In Fig.~\ref{fig:Robustness}, the effects on the giant component of removing nodes in targeted  and random manners are illustrated.

\begin{figure}[t!]
\begin{center}
\includegraphics[width=0.9\columnwidth, angle=0]{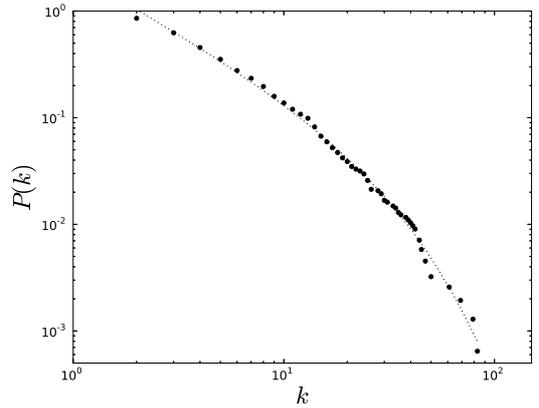}
\caption{The cumulative degree distribution for the amalgamation of all 18 data sets. The dotted line is a truncated power-law fit after Eq.(\ref{eqn:truncated}).}
\label{fig:degree_all}
\end{center}
\end{figure}

The degree distribution for the entire saga society analysed (comprising all 18 narratives) is displayed in Fig.~\ref{fig:degree_all} with a best fit to Eq.(\ref{eqn:truncated}). 
One finds $\hat{\gamma} = 2.1 \pm 0.2$,  $\kappa = 26 \pm 2$.

Finally using the same Girvan-Newman algorithm to break the network down into components, and evaluating the modularity at each interval, we find $Q$ reaches a plateau at over 0.7 with $n=9$ communities. As there are 18 separate sagas, this indicates that they are not easily split back down into their individual components. 
However a number of the tales contain only about 20 or 30 characters, some whom appear in more than one narrative. The difficulty in separating them reflects the inter-connectedness of the {\emph{{\'{I}}slendinga s{\"{o}}gur}}.

%%%%%%%%%%%%%%%%%%%%%%%%%%%%%%%%%%%%%%%%%%%%%%%%%%%%%%%%%%%%%%%%%%%%%%%%%%%%%%%%%%%%%%%%%%
\subsection{Comparisons to other networks}
%%%%%%%%%%%%%%%%%%%%%%%%%%%%%%%%%%%%%%%%%%%%%%%%%%%%%%%%%%%%%%%%%%%%%%%%%%%%%%%%%%%%%%%%%%

The five individual saga networks have many properties of real social networks -- they are small world, have high clustering coefficients,  are structurally balanced and contain sub-communities. Most are well described by power-law degree distributions, and the family sagas, in particular, have non-negative assortativity.

In Ref.~\cite{EPL}, we  studied the properties of networks associated with three epics, the {\emph{Iliad}} from ancient Greece, the Anglo-Saxon {\emph{Beowulf}} and the Irish {\emph{T{\'{a}}in B{\'{o}} Cuailnge}}. Of these, we found that the {\emph{Iliad}} friendly network has all the above properties. 
Although {\emph{Beowulf}} and the {\emph{T{\'{a}}in}} also have many of these features, they are notable in that their  friendly and full networks are disassortative. 
{\emph{G{\'{i}}sla saga S{\'{u}}rssonar}} is also dissassortative implying that its network is more similar to these two heroic epics rather than the other family sagas.

Conflict is an important element of the three narratives analysed in Ref.~\cite{EPL}, in that hostile links are generally formed when characters who were not acquainted meet on the battlefield.  
This is quite different for the  {\emph{{\'{I}}slendinga s{\"{o}}gur}}, for which many hostile links are due to blood feuds as opposed to  armies at war. 
Here hostile links are often formed between characters who are already acquainted.
For this reason, there is little difference between the properties of the full network and its positive sub-network, as indicated in Table~\ref{tab:Networks2}.

In comparison to the three epics analysed in Ref.~\cite{EPL}, the  {\emph{{\'{I}}slendinga s{\"{o}}gur}} are most similar of the \emph{Iliad} friendly network in that they are both small world, assortative and their degree distributions follow power laws with exponential cut-offs. However, the amalgamated saga networks differs from  the \emph{Iliad} in that the overall network of the latter is disassortative, a difference reflecting the nature of conflict in the stories. 

To summarise, network analysis indicates that the  {\emph{{\'{I}}slendinga s{\"{o}}gur}} comprise a highly interlinked set of narratives, the structural properties of which  are not immediately distinguishable to those of real social networks.

%..................................................................................
%..................................................................................

%%%%%%%%%%%%%%%%%%%%%%%%%%%%%%%%%%%%%%%%%%%%%%%%%%%%%%%%%%%%%%%%%%%%%%%%%%%%%%%%%%%%%%%%%%
\section{Conclusions}
\label{5}
%%%%%%%%%%%%%%%%%%%%%%%%%%%%%%%%%%%%%%%%%%%%%%%%%%%%%%%%%%%%%%%%%%%%%%%%%%%%%%%%%%%%%%%%%%

We have analysed the networks of the sagas of  Icelanders and compared them to each other, to real and fictitious 
social networks and to the networks underlying  other European epics.

Of the five narratives with the largest cast, {\emph{Laxd{\ae}la saga}} and {\emph{G{\'{i}}sla saga S{\'{u}}rssonar}}  appear most dissimilar  to each other. 
{\emph{Laxd{\ae}la saga}} has an exponential degree distribution possibly indicating that the higher degree characters are less important in terms of the overall properties of the network 
as compared to the others.
Indeed, {\emph{Laxd{\ae}la saga}}  is strongly assortative.
{\emph{G{\'{i}}sla saga S{\'{u}}rssonar}} on the other hand is strongly disassortative indicating that protagonist dominates the properties of the network. {\emph{Egils saga Skallagr{\'{i}}mssonar}} is similar to {\emph{Nj{\'{a}}ls saga}} and {\emph{Vatnsd{\ae}la saga}}, however it too has distinguishing features. Despite these differences, there are many properties common to the sagas' social networks.

Amalgamating the five major sagas generates a small-world network with a power-law degree distribution and an exponential cut-off which is assortative and contains strong community structure.
This amalgamated network can mostly be decomposed using the algorithm of Girvan-Newman~\cite{GirvanNewman}.
In this case, two saga networks, namely those of {\emph{Laxd{\ae}la saga}} and  {\emph{Nj{\'{a}}ls saga}}, emerge with a large degree of overlap.
An eventual separation of these two sagas is only achieved by also splitting the {\emph{Laxd{\ae}la}} network into two. \emph{Laxd{\ae}la saga} is also easily fragmented using the Girvan-Newman algorithm  indicating that it seems to consist of weakly connected sub-components some of which overlap with {\emph{Nj{\'{a}}ls saga}}.
  
%..................................................................................
%..................................................................................

The further amalgamation of the five sagas with 13 other narratives was then analysed.
The resulting saga society is similar to the family-based networks of {\emph{Nj{\'{a}}ls saga}}, {\emph{Egils saga}} and {\emph{Vatnsd{\ae}la saga}}, though it has a higher assortativity.
Hence the full society generated from these texts has similar properties to those of many real social networks, in that they are small world,
structurally balanced, follow a power-law degree distribution with an exponential cut-off and are assortative by degree. It is also not easy to break it back down to it's individual 18 components.

The {\emph{{\'{I}}slendinga s{\"{o}}gur}} hold a unique place in world literature and have fascinated scholars throughout the generations. 
Instead of analysing the characters themselves, we provide information on how characters are interconnected to compare the social structures underlying the narratives. In this way, we provide a novel approach to compare sagas to each other and to other epic literature, identifying similarities and differences between them.
The comparisons we make here are from a network-theoretic point of view and more holistic information from other fields (such as comparative mythology and archaeology) is required to inform further.
In a similar spirit to Ref.~\cite{Dunbar}, we also conclude that whether the sagas are historically accurate or not,
the properties of the social worlds they record are similar to those of real social networks.
Although one cannot conclusively determine whether the saga societies are real, on the basis of network theory, we can conclude 
that they are realistic.

\vspace{0.5cm}
\noindent
{\bf{Acknowledgements:}}
We wish to thank Thierry Platini and Heather O'Donoghue for helpful discussions and feedback. We also thank Joseph Yose and Robin de Regt for help gathering data.
This work is supported by The Leverhulme Trust under grant number F/00~732/I and in part by  a Marie Curie IRSES grant within the 7th European Community Framework Programme.

\newpage

%%%%%%%%%%%%%%%%%%%%%%%%%%%%%%%%%%%%%%%%%%%%%%%%%%%%%%%%%%%%%%%%%%%%%%%%%%%%%%%%%%%%%%%%%%

%%%%%%%%%%%%%%%%%%%%%%%%%%%%%%%%%%%%%%%%%%%%%%%%%%%%%%%%%%%%%%%%%%%%%%%%%%%%%%%%%%%%%%%%%
\appendix*
\section{{\emph{{\'{I}}slendinga S{\"{o}}gur}}}
%%%%%%%%%%%%%%%%%%%%%%%%%%%%%%%%%%%%%%%%%%%%%%%%%%%%%%%%%%%%%%%%%%%%%%%%%%%%%%%%%%%%%%%%%

The following is the full list of translations of sagas used to construct the amalgamated network of 18 narratives.
The first 17 saga versions are taken from the translations of Ref.~\cite{Smiley2000} and the edition of the last saga is that contained in  Ref.~\cite{Njal}.\\

\noindent Egil's Saga \\
The Saga of the People of Vatnsdal \\
The Saga of the People of Laxardal \\
Bolli Bollason's Tale \\
The Saga of Hrafnkel Frey's Godi \\
The Saga of the Confederates \\
Gisli Sursson's Saga \\
The Saga of Gunnlaug Serpent Tongue \\
The Saga of Ref the Sly \\
The Saga of the Greenlanders \\
Eirik the Red's Saga \\
The Tale of Thorstein Staff-Struck \\
The Tale of Halldor Snorrason II \\
The Tale of Sarcastic Halli \\
The Tale of Thorstein Shiver \\
The Tale of Audun from the West Fjords \\
The Tale of the Story-wise Icelander \\
Njal's Saga

\end{document}